# A simple model for elastic and viscoelastic punch indentation problems with experimental validation

A. SAMEUR[1], H. P. YIN[1], D. DUHAMEL[1], V. VILKE[2]

[1] Université Paris-Est, Institut Navier, LAMI, Ecole Nationale des Ponts et Chaussées, 6 et 8 Avenue Blaise Pascal, Cité Descartes, Champs sur Marne, 77455 Marne La Vallée, FRANCE

[2] Moscow State University M.V. Lomonosov, Faculty of Mechanics and Mathematics, Main Building, MSU, Vorobjovy Gory, Moscow, 119899, RUSSIA

Contact author:

Duhamel Denis

LAMI, ENPC-LCPC, Institut Navier

Ecole Nationale des Ponts et Chaussées

6 et 8 Avenue Blaise Pascal, Cité Descartes, Champs sur Marne

77455 Marne La Vallée cedex 2, FRANCE

Tel : 00 33 1 64 15 37 28

Fax : 00 33 1 64 15 37 41

Email : duhamel@lami.enpc.fr

# ABSTRACT

This paper presents an analytical model of punctual elastic contact between a rigid body of arbitrary geometry and a plane surface. A simple analytical model is developed in order to evaluate the contact force knowing the volume of interpenetration, the surface and the perimeter of the base of this volume and the mechanical characteristics of surfaces in contact. Analytical and experimental validations are made for this model in the case of simple shapes (spherical, conical and pyramidal). Next, an approach for the resolution in case of contact between a rigid body and a viscoelastic plane is presented. The elastic constants are replaced by an integral operator corresponding to the viscoelastic stress-strain relation. At last, the viscoelastic punctual contact is studied analytically and validated experimentally.

# 1. Introduction

The study of contact problems has its origins in the last quarter of the 19th century when Hertz [6] and Boussinesq [1] presented solutions to the contact of linear elastic materials. Several solutions were derived from the solution of Boussinesq, an excellent account of which is given in Galin's book [4] and in Johnson [8]. Later, Sneddon [14] established a solution of the axisymmetric Boussinesq problem which enabled him to deduce simple formulas giving the penetration $\delta$ of a punch of arbitrary profile as well as the total force necessary to ensure this penetration.

Then the interest of the viscoelastic contact problem emerged, Graham [5] gave an expression for the displacement stress field produced at any point of a linear homogeneous and isotropic viscoelastic half space by an arbitrary time dependent distribution of pressure acting on its boundary. The distribution of normal surface tractions prevailing when a rigid punch of arbitrary profile is pressed against the surface of a viscoelastic half space is determined in terms of a one parameter family of solutions to the corresponding elastic problem. One of the approaches for the resolution of a contact problem of a rigid sphere with a viscoelastic material was suggested by Radok [13]. It replaces elastic constants by an integral operator corresponding to the viscoelastic stress strain relation in which the radius of the contact area is a monotonically increasing function of time. Later, Hunter [7] studied the Hertz's problem in the case of the rebound of a rigid sphere on a viscoelastic half space so that the contact radius increases monotonically to a maximum and then decreases to zero. It was assumed that the distribution of pressure remains elliptic as in the elastic case. Ting [15] presented a method so that the problem of contact could be solved for an arbitrary contact radius.

Numerical methods for the resolution of the contact were also developed. Webster & Sayle [17] and Chang & Gao [3] developed a numerical model for the elastic contact of rough surfaces. Younguing & Linquing [18] proposed a numerical modelling for an elastic *3D* contact of rough surfaces. The authors conclude that the interaction of asperities has a great effect on the calculated results, particularly on the contact deformation. The finite element method was used by Kucharski et al [10] who modelled the contact between a sphere and a rigid plane. They obtained a relation between the weight and the surface of contact. Kane et al [9] developed robust contact algorithms able to deal with complex contact situations involving several bodies with corners. However these numerical methods lead to time consuming computations.

This paper presents a semi analytical model which uses a theory of the interaction potential between a rigid body and an elastic or viscoelatic solid. The aim is to get simple and very fast estimates for the contact force for given body shapes. The contact force can be expressed in terms of the volume of interpenetration, the surface and the perimeter of the base of this volume as well as the mechanical characteristics of surfaces in contact. Then, an approach for the resolution in the case of a rubber block with a viscoelastic behaviour is presented. Radok's approach [13] is used; it replaces elastic constants by an integral operator corresponding to the viscoelastic stress-strain relation. Experimental validations are made for different shapes of simple rigid bodies (spherical, conical and pyramidal). Finally, a comparison of contact forces in the elastic and viscoelastic case is given.

## 2. Classical contact theory

One of the first studies concerning the evaluation of the contact tensions between elastic solids was proposed by Hertz [6]. To express the normal force $P$ versus the indentation, he made the assumptions that the sector of contact is elliptic, that each solid can be regarded as an elastic half space, that there is no friction between the two surfaces in contact and that the surfaces are continuous and no conform. These assumptions lead to the following well known relation

$$P = K \delta^{\frac{3}{2}} \quad (1)$$

with

$$K = \frac{4E^* \sqrt{R}}{3} \quad (2)$$

where $P$ is the normal force applied, $R$ is the equivalent radius of curvature and $E^*$ represents the Young's equivalent modulus. The theory of the Hertz's contact is limited to surfaces with smooth and continuous profiles with finite forces everywhere. The problem is different for a surface having edges or corners. The deformation must be sufficiently small in order to be in the field of the linear theory of elasticity.

Consider a cone in contact with a plane surface, the area of contact is supposed to be small compared to the size of the two solids. The deformation is shown in Fig.1 which presents also the pressure distribution in the contact zone. If the smooth sides of the cone are prolonged beyond the contact surface, the pressures must equal zero at the edges. A classical approach to find the efforts in an elastic half space due to external tractions is given by Boussinesq [1] and Cerruti [2], who used the potential theory. Love [12] applied the classical approach of the potential function to the cone and evaluated the contact pressure by the formula

$$p(s) = \frac{1}{2} E^* \cot \alpha \cosh^{-1}(a/s) \quad (3)$$

where $s$ is the distance between the origin $O$ and a point ranging between $O$ and $a$. Thus the normal force is given by

$$P = \frac{1}{2} \pi a^2 E^* \cot \alpha \quad (4)$$

The quantities $\alpha$ and $a$ are defined in Fig. 1.

## 3. Contact between two bodies of complex shape

### 3.1. *Interaction potential theory*

Suppose that two bodies $b_1$ and $b_2$ are in contact at the point $O$. $Oxy$ is the common tangent plane with axis $OZ$ directed towards the body $b_1$ (see Fig. 2.a). The distance between the points $M_1$ and $M_2$ of the two bodies in the vicinity of the point $O$ equals

$$z_1 - z_2 \approx Ax^2 + By^2 \tag{5}$$

The points $M_1$ and $M_2$ have coordinates $(x, y, z_1)$ and $(x, y, z_2)$ and belong to the surfaces of the bodies. The values A and B are constant. In general we find an additional term $2Cxy$ in the last part of expression (5) but it can be eliminated by a suitable choice for the axes $Ox$ and $Oy$. If the body $b_1$ moves by a distance $\delta$ in the negative direction of axis $Oz$ and if it is assumed that the bodies interpenetrate without deformation, there will be an intersecting domain $\Gamma$ whose projection on the $Oxy$ plan forms the surface (see Fig. 2b)

$$\sigma = \{x,y : Ax^2 + By^2 \leq \delta\} \tag{6}$$

with the boundary

$$\partial\sigma = \{x, y : Ax^2 + By^2 = \delta\} \tag{7}$$

The assumption that the bodies interpenetrate without deformation is an approximation. But this is not too far from reality and this will allow getting the simple expressions that follow. It will also be justified by comparison with the experimental results.

Three characteristics of the intersecting domain will play an important role, namely, the volume $V$ of $\Gamma$, the surface $S$ and the perimeter $p$ of $\sigma$

$$V = \iint_\sigma (\delta - Ax^2 - By^2)\,dx\,dy = \frac{\pi \delta^2}{2\sqrt{AB}} \tag{8}$$

$$S = \iint_{\sigma} dx\,dy = \frac{\pi \delta}{\sqrt{AB}}, \tag{9}$$

$$p = \left(\frac{16\delta}{A}\right)^{1/2} E_1(\sqrt{A/B}) \tag{10}$$

where $E_1(\sqrt{A/B})$ is the elliptic integral of second species given by

$$E_1(\sqrt{A/B}) = \int_0^{\pi/2} (1 - (1 - A/B)\sin^2 \varphi)^{1/2} d\varphi \tag{11}$$

According to the traditional results of Hertz, in the case where the contact zone is an ellipse of surface $\sigma' = \left\{ x,y : \frac{x^2}{a^2} + \frac{y^2}{b^2} \leq 1 \right\}$, the force $P$ and the coefficients of surfaces are linked by the following relations Love [11]

$$A = \frac{3P(\theta_1 + \theta_2)}{2a^3} I_1(k), \quad I_1(k) = \int_0^\infty \frac{dz}{\sqrt{(1+z^2)(1+k^2 z^2)^3}}$$

$$B = \frac{3P(\theta_1 + \theta_2)}{2b^3} I_2(k), \quad I_2(k) = \int_0^\infty \frac{dz}{\sqrt{(1+z^2)(1+k^{-2} z^2)^3}} \tag{12}$$

$$\delta = \frac{3P(\theta_1 + \theta_2)}{2a} F_1(k), \quad F_1(k) = F(k_1) = \int_0^{\frac{\pi}{2}} \frac{d\varphi}{\sqrt{1 - k_1^2 \sin^2 \varphi}}$$

where $k = \frac{b}{a} < 1$, $k_1^2 = 1 - k^2$, $\theta_i = \frac{1 - v_i^2}{\pi E_i}$, $i = 1,2$, $F(k_1)$ is the elliptic integral of first species, $E_i$ and $v_i$ are the Young's modulus and the Poisson's ratio of the body $b_i$. A relation between $A$ and $B$ can be obtained from Eq. (12)

$$\frac{A}{B} = \frac{k^3 I_1(k)}{I_2(k)} \tag{13}$$

It follows that the relation $b/a = k < 1$, which does not depend on the load $P$, is defined by the geometrical characteristics of surfaces in contact. The force $P$ is given by (see Appendix A.)

$$P = \frac{2(I_1(k)I_2(k))^{1/4} \delta^{3/2}}{3(\theta_1 + \theta_2) k^{3/4} (F_1(k))^{3/2} (AB)^{1/4}} \tag{14}$$

The potential $U$ for the interaction of the bodies in contact, is calculated from the expression of virtual works by

$$\delta U = P \partial \delta \Rightarrow U(\delta) = \frac{4(I_1(k)I_2(k))^{1/4} \delta^{5/2}}{15(\theta_1 + \theta_2) k^{3/4} (F_1(k))^{3/2} (AB)^{1/4}} \tag{15}$$

Let us represent the relation (15) by the form

$$U = f(k) V^\beta S^\gamma p^\xi \tag{16}$$

where $V$, $S$ and $p$ are defined in the relations (8), (9) and (10). By comparing the powers of $\delta$ and $(AB)$ in the expressions (15) and (16) of $U$, we find $\beta = 2$ and $\gamma = -\frac{(3+\xi)}{2}$. This leads to

$$f(k) = \frac{4^{2-\xi} f_1(k)}{15 \pi^{(1-\xi)/2} (\theta_1 + \theta_2)} \tag{17}$$

where

$$f_1(k) = \frac{I_1^{(1+\xi)/4} I_2^{(1-\xi)/4}}{k^{3(1-\xi)/4} F_1(k)^{3/2} E_1^\xi(\sqrt{A/B})} \tag{18}$$

To study the function $f_1(k)$, let us express $I_1(k)$ and $I_2(k)$ by the elliptic integrals

$$I_1(k) = \frac{F_1(k) - E_1(k)}{1 - k^2} \tag{19}$$

$$I_2(k) = \frac{k}{1 - k^2} (E_1(k) - k^2 F_1(k)) \tag{20}$$

Finally we obtain

$$f_1(k) = \frac{(1 - E_1 F_1^{-1})^{(1+\xi)/4} (E_1 F_1^{-1} - k^2)^{(1-\xi)/4}}{k^{(1-\xi)/2} (1-k^2)^{1/2} F_1 E_1^\xi(\sqrt{A/B})} \tag{21}$$

The function $f_1(k)$ remains bounded if $k$ tends towards $1$, since according to Wittaker [16]

$$\lim_{k \to 1} I_1(k) = \frac{\pi}{4} \text{ and } \lim_{k \to 1} F_1(k) = \frac{\pi}{2} \tag{22}$$

$$\lim_{k \to 1} \sqrt{A/B} = 1 \text{ and } E_1(1) = \frac{\pi}{2} \tag{23}$$

Then it follows

$$\lim_{k \to 1} f_1(k) = \frac{2^{\xi+0.5}}{\pi^{1+\xi}} \tag{24}$$

The numerical analysis of the function (21) represented in Fig. 3 shows us that the function $f_1(k)$ is nearly constant for $\xi = 0.5$ with the average value $0.36$. Let us express the potential $U$ by the formula

$$U = c \frac{8}{15\pi^{1/4}(\theta_1 + \theta_2)} \frac{V^2 p^{1/2}}{S^{7/4}} \tag{25}$$

The parameter $c$ is a constant which depends only on the geometry of contact surfaces. According to the preceding development $c$ equals 0.36 for regular contact surfaces and for $\xi = 0.5$. It is supposed that the potential can be written in the same form (25) in the case of a contact of irregular surfaces. We will identify analytically and by experiments the constant $c$ in order to validate this generalized formula in the case of bodies with arbitrary shapes.

3.2. *Validation in simple shape cases*

For this aim, three surface shapes are considered: the contact is between a plane surface and with spherical, conical and pyramidal rigid surfaces. For each case the relation between the force and the interpenetration will be determined. The constant $c$ is then identified from experimental tests.

To validate and generalize the expression of the potential (25) for any surface shape, let's take the example of a spherical form (see Fig. 4a). Once the expressions of the volume, the surface and the perimeter are known, we can replace then in Eq. (25). The expressions of the volume, the surface and the perimeter of the part in contact are given by

$$V = \frac{\pi}{6}\delta(3r^2 + \delta^2)$$

$$S = \pi r^2 \qquad (26)$$

$$p = 2\pi r$$

where $r$ is the radius of the base of the volume in contact and $\delta$ is the interpenetration of the sphere in the plane surface. They are expressed as function of the angle $\alpha$ by the relations

$$r = R\sin\alpha \quad \text{and} \quad \delta = R(1-\cos\alpha) \qquad (27)$$

Suppose that $\alpha$ is very small thus

$$r = R\alpha \quad \text{and} \quad \delta = R\frac{\alpha^2}{2} \qquad (28)$$

The relation between $r$ and $\delta$ is found by $r^2 = 2R\delta$. It is assumed that $r << R$, thus $\delta^2 << r^2$. So the potential is summarized in the case of a plane contact/sphere by

$$U(\delta) = c_s \frac{4}{15}\pi^{3/2} E^* \sqrt{R}\ \delta^{5/2} \qquad (29)$$

The force $P$ then equals

$$P = \frac{\partial U}{\partial \delta} = c_s \frac{2}{3}\pi^{3/2} E^* \sqrt{R}\ \delta^{3/2} \qquad (30)$$

The relation (30) expresses the force as function of the interpenetration in the case of a contact between a plane surface and a spherical surface. The constant $c_s$ can be determined by identification with the Hertz's law which leads to $c_s = \frac{2}{\pi^{3/2}} = 0.36$.

It would also be useful to know the expression of the force in case of a contact between a plane surface and a cone as well as in the case of a contact between a plane surface and a pyramid. An approximation of $c$ for various surfaces in contact enables one to generalize the contact law (25).

The same method is used for a contact between a conical surface with angle $\alpha$ and a plane surface (see Fig. 4b). The expressions of the volume, the surface and the perimeter of contact are evaluated and replaced in the expression of the potential (25). The volume of the part in contact is evaluated by

$$V = \frac{\pi}{3} r^2 \delta \qquad (31)$$

The surface of the base in contact is $S = \pi r^2$, and the perimeter of the surface $S$ is $p = 2\pi r$.

The relation between $r$ and $\delta$ is known to be $tg\alpha = \frac{r}{\delta}$. So the potential equals

$$U = c_c \frac{8}{15} \frac{\sqrt{2}}{9} E^* \pi^{3/2} tg\alpha \, \delta^3 \qquad (32)$$

The force $P$ is written as

$$P = c_c \frac{8\sqrt{2}}{45} \pi^{3/2} E^* tg\alpha \, \delta^2 \qquad (33)$$

In order to find the relation between the interpenetration $\delta$ and the radius $a$, let us take the Boussinesq's equations for a point force on a half space. It follows that the relation of the interpenetration $\delta$ can be written under the form

$$\delta(s) = \frac{p}{4\pi G} \left( \frac{2(1-\upsilon)}{s} \right) \qquad (34)$$

where $G = \frac{E}{2(1+\upsilon)}$ is the shear modulus and $\upsilon$ is the Poisson's ratio. Knowing the pressure repartition (3) given by Love [12], the interpenetration $\delta$ which is the displacement at the apex of the cone, is written as

$$\delta = \frac{1}{4\pi G} \int_0^a 2\pi s \, p(s) \left( \frac{2(1-\upsilon)}{s} \right) ds \qquad (35)$$

It becomes

$$\delta = \frac{\pi}{2} a \cot \alpha \frac{(1-\upsilon) E^*}{2G} \qquad (36)$$

So the relation (4) can be written in the form

$$P = \frac{8}{\pi} \frac{G^2}{(1-\upsilon)^2 E^*} tg\alpha \, \delta^2 . \qquad (37)$$

If the assumption is made that the Young's modulus of the body (1) is very small in front of the Young's modulus of the body (2) $E_1 = E << E_2$, we then obtain

$$\frac{1}{E^*} \approx \frac{(1-\upsilon_1^2)}{E_1} = \frac{(1-\upsilon^2)}{E} \qquad (38)$$

It follows that $G = \frac{(1-\upsilon)}{2} E^*$, and the relation (36) becomes

$$P = \frac{2}{\pi} E^* tg\alpha \, \delta^2 \qquad (39)$$

By comparison with relation (33) we find $c_c = \frac{45}{4\sqrt{2}\,\pi^{5/2}} \approx 0.45$.

3.3. *Application to a pyramid*

After considering the case of a contact between a plane surface with a sphere and a cone, it would be interesting to study the case of a contact between a plane surface and a pyramid (see Fig. 4c). The volume of the part in contact is evaluated by

$$V = \frac{1}{3} r^2 \delta \qquad (40)$$

The surface of the base in contact is $S = r^2$ and the perimeter of the surface $S$ is $p = 4\,r$.

The relation between $r$ and $\delta$ is $tg\alpha = \frac{\sqrt{2}}{2} \frac{r}{\delta}$. So the potential equals

$$U = c_p \frac{8}{15} \frac{2\sqrt{2}}{9} E^* \pi^{3/4} tg\alpha \, \delta^3 \qquad (41)$$

The force is thus written in the form

$$P = \frac{\partial U}{\partial \delta} = c_p \frac{16\sqrt{2}}{45} E^* \pi^{3/4} tg\alpha \, \delta^2 \tag{42}$$

A model of the elastic contact between a plane surface and various surface shapes has been presented. The force $P$ is expressed according to the interpenetration $\delta$, the geometrical and mechanical characteristics of the bodies in contact, and also by a constant $c$ which was analytically identified for the spherical and conical case. The constants $c_s$, $c_c$ and $c_p$ will also be identified experimentally and the expressions (30), (39), (42) will so be validated.

**4. Extension to materials with a viscoelastic behaviour**

Let us consider a pure shear stress, the stress strain relation expressed according to the shear modulus is $s = 2Ge$. One of the approaches for the resolution of a contact problem between a rigid sphere and a viscoelastic material was suggested by Radok [13]. His approach replaces elastic constants by an integral operator corresponding to the viscoelastic stress strain relation

$$s(t) = \int_0^t \psi(t-t') \frac{\partial e(t')}{\partial t'} dt' \tag{43}$$

Let us suppose that the variation of the force $P$ according to the interpenetration $\delta$ in the case of elastic contact is written as

$$P = 2G \, Q \, \delta^\gamma \tag{44}$$

where $\gamma$ and $Q$ depend on the form of the contact surface. By applying the Radok's method [13], the expression of the force becomes

$$P(t) = Q \int_0^t \psi(t-t') \frac{d}{dt'} \delta^\gamma(t') dt' \tag{45}$$

By taking into account the parameters $\gamma$ and $Q$ defined in Table 1, the expression of the force is finally given by

- for a contact between a rigid sphere and a viscoelastic plane

$$P(t) = \frac{8}{3}\sqrt{R}\int_0^t \psi(t-t')\frac{d}{dt'}\delta^{3/2}(t')dt' \tag{46}$$

- for a contact between a rigid cone and a viscoelastic plane

$$P(t) = \frac{4}{\pi} tg\,\alpha \int_0^t \psi(t-t')\frac{d}{dt'}\delta^2(t')dt'. \tag{47}$$

- for a contact between a rigid pyramid and a viscoelastic plane

$$P(t) = c_p \frac{32}{45}\sqrt{2}\pi^{3/4}\cot\alpha \int_0^t \psi(t-t')\frac{d}{dt'}\delta^2(t')dt' \tag{48}$$

**5. Experimental validation for elastic contacts**

In this purpose an experimental device is established. A compression *Instron* machine, bodies with various shapes (spherical, conical and pyramidal), *LabView* software for the acquisition of the results and a rubber block are used. Experimental relations between the force and the interpenetration are established for each body in contact with the rubber block. First the Young's modulus of the rubber block is identified by a relaxation test. Knowing the characteristics of the rubber block and the contact bodies, the coefficients $c_s$, $c_c$, $c_p$ corresponding to the contact models for the spherical, conical and pyramidal cases will be identified.

5.1. *Experimental identification of viscoelastic parameters*

A simple compression test on the rubber block presented in Fig. 5 is abruptly and quickly carried out by using a compression machine. A deformation of 8% is maintained constant and the evolution of the stress versus the time is recorded. It is noticed an abrupt and rapid increase of the stress to $\sigma_0 = 0.93$ Mpa just after the load. This deformation is maintained by blocking the crossbar for a long time until the stabilization of the stress which tends towards $\sigma_\infty = 0.6\,MPa$. The stress versus the time is written under the form of a Prony's series

$$\sigma(t) = (A_n e^{-t/\tau_n} + A_{n-1} e^{-t/\tau_{n-1}} + ... + A_1 e^{-t/\tau_1} + E_\infty)\varepsilon_0 \tag{49}$$

For the rubber block four characteristics times are found and the results are represented in table 2. After reconstitution of the relaxation curve from characteristics times and amplitudes, in Fig. 6, we notice that there is a good concordance of the two curves. Using less than four characteristic times doesn't allow a good reconstitution of the curve over the whole time range. The static Young's modulus of the block is found to be $E_\infty = \dfrac{\sigma_\infty}{\varepsilon_0} = 7.5\,MPa$

5.2. *Experimental validation of the elastic contact model*

5.2.1. *Contact between a rubber block and a sphere*

The experimental device is the same as the one used for the relaxation test (see Fig. 7a). The load is carried out at a speed of 0.001 *mm/s* to simulate a static loading with the Young's modulus $E_\infty = 7.5\,MPa$. Fig. 8a represents the evolution of the force versus the interpenetration. By linear regression of the curve we obtained a line that has the following equation $Ln(P) = 1.51 Ln(\delta) + 14.10$. According to Eq. (30) and to the experimental results of the contact between the rubber block and the steel ball, it was found

$$c_s \frac{2}{3}\pi^{3/2} E^* \sqrt{R} = e^{14.10} \tag{50}$$

Knowing the mechanical and geometrical characteristics of the two bodies in contact, the experimental value from the coefficient $c_s$ is deduced and equals 0.34. The theoretical value of the coefficient $c_s$ equals 0.36, which is slightly different from the experimental value. This result allows to validate the experiment by recovering the Hertz's result.

*5.2.2. Contact between a rubber block and a cone*

The same test as the one used for the contact between the rubber block and the sphere is carried out. Two steel cones that have different angles ($\alpha = 60°$ and $\alpha = 45°$) are used in order to study the influence of the angle on the distribution of the force. The experimental device is presented in Fig. 7b. The force versus the interpenetration for each cone in contact is given in Fig. 8b. By plotting the logarithmic curve of the force as function of the interpenetration and after having carried out a linear regression for each curves it is found for the cone with $\alpha = 60°$ that $Ln(P) = 1.99\,Ln(\delta) + 16.15$ and for the cone with $\alpha = 45°$ that $Ln(P) = 2.00\,Ln(\delta) + 15.67$.

The evaluation of the force from the expression (33) adapted for the case of a contact between a plane surface and a cone is given by

$$P = \frac{\partial U}{\partial \delta} = c_c \frac{8\sqrt{2}}{45} \pi^{3/2} E^* tg\alpha\, \delta^2 \qquad (51)$$

The values of the coefficients are deducted and $c_c = 0.43$ for an angle $\alpha = 60°$ and $c_c = 0.45$ for an angle $\alpha = 45°$. The value of the coefficient $c_c$ does not differ much for different angles of the cone and from the analytical value which equals 0.45.

*5.2.3. Contact between a rubber block and a pyramid*

The expression of the force versus the interpenetration is

$$P = \frac{\partial U}{\partial \delta} = c_p \frac{16\sqrt{2}}{45} E^* \pi^{3/4} tg\alpha\, \delta^2 \qquad (52)$$

In order to identify the coefficient $c_p$, the experimental device used for the identification of the coefficients $c_s$ and $c_c$ is taken again. The pyramid has an angle $\alpha = 68°$ (see Fig. 7c.). After having plotted the logarithmic curve (see Fig. 8c.) of the force variation versus the interpenetration, one obtains

$$c_p \frac{16\sqrt{2}}{45} E^* \pi^{3/4} tg\alpha = e^{16.53} \quad (53)$$

Which leads to $c_p = 0.53$.

The coefficients $c_s$ and $c_c$ were identified experimentally and analytically. $c_p$ was identified only experimentally since there is no analytical model for the pyramidal case. In Table 3 the values obtained for various surfaces are presented. The differences between the experimental and the analytical results are quite small. Let's conclude that the value of the coefficient $c_s$ equals 0.36. For the conical case the value of the coefficient $c_c$ equals 0.45, which is true for any angle α. Considering the absence of a classical analytical theory for the pyramidal shape, we can however conclude that the contact law has the same shape as for the other cases with $c_p$ which equals 0.53.

## 6. Experimental validation for viscoelastic contacts

The expression of the force according to the interpenetration is put in the form (46), (47) and (48). In the same way as for the case of an elastic contact, experimental tests are carried out in order to validate viscoelastic contact models.

### 6.1. *Contact between a rubber block and a steel ball*

In this test the same experimental device is used, except that instead of charging at a speed of 0,001 *mm/s*, the loading is done at a speed of 5 *mm/s*. The same rubber block and the same steel ball are used so the geometrical and mechanical characteristics are well known. The test does not last more than one second so only the characteristic time $\tau_4 = 3,37s$ is taken into account. From Eq. (49) one obtains

$$\sigma(t) = (A_4 e^{-t/\tau_4} + E') \varepsilon_0 \quad (54)$$

with $E' = E_\infty + A_1 + A_2 + A_3$. In order to solve Eq. (346), let's put $\delta(t') = V t'$. The shear modulus for an incompressible material is $G = \dfrac{E}{3}$. It results that Eq. (46) gives

$$P(t) = \frac{8}{3}\sqrt{R}\int_0^t \frac{2}{3}\left(A_4\, e^{-(t-t')/\tau_4} + E'\right)\frac{d}{dt'}(V t')^{3/2}\, dt' \tag{55}$$

Which after integration leads to

$$P(t) = \frac{8}{3}\sqrt{R}\, V^{3/2}\left[\frac{2}{3} E' t^{3/2} + A_4\, \tau_4\, t^{1/2} + \frac{i\sqrt{\pi}}{2}\tau_4^{3/2}\, A_4\, e^{-t/\tau_4}\, erf\left(i\sqrt{\frac{t}{\tau_4}}\right)\right] \tag{56}$$

where *erf* is the error function given by

$$erf(i x) = \frac{-2}{i\sqrt{\pi}}\int_0^x e^{y^2}\, dy \tag{57}$$

In Fig. 9, we can see the evolution of the force $P$ according to the interpenetration $\delta$ for the spherical, conical and pyramidal cases. The curves show that in all cases there is a good concordance between the two results. A comparison between the results obtained in the case of the elastic analytic model and in the case of the viscoelastic analytic model concludes that the value of the force in the case of a viscoelastic contact is higher by 24% than the value of the force in the case of an elastic contact.

## 7. Conclusion

The theoretical model of the potential interaction for contacts between regular surfaces was presented and generalized for irregular surfaces. The validation of this model in the case of a contact between a plane surface with a sphere, a cone and a pyramid was made. Then an approach of the solution of a viscoelastic contact problem was seen, where the method of Radok [13], which consists of replacing elastic constants by an integral operator, was applied. The viscoelastic behaviour of the material was modeled by a Prony's series. The Young's modulus and the characteristic times of a rubber block were identified. Once these characteristics were identified, tests were carried out in which the rubber block was put into contact with various surfaces (spherical, conical and pyramidal). These tests made it possible to identify the coefficients $c_s$, $c_c$ and $c_p$. Thus, we validated the models of elastic contact.

Using the same method, tests of viscoelastic contact at a speed of 5 *mm/s* with the same rubber block and the various bodies were carried out. The results were compared to the results of a viscoelastic contact model. The results of the model and those of the experiments agree very well. A comparison between the results of the elastic contact model and the results of the viscoelastic contact model shows that for a given value of interpenetration $\delta$, the viscoelastic force is 25% higher than the elastic force for this load speed. All these results validate the simple semi-analytical model developed in this paper for elastic and viscoelastic contacts of bodies with arbitrary shapes.

**Appendix A. Calculation of the force *P***

The force *P* and the coefficients of surfaces $\sigma$ and $\sigma'$ are linked by the following relations Love [11]

$$A = \frac{3P(\theta_1 + \theta_2)}{2a^3} I_1(k), \quad I_1(k) = \int_0^\infty \frac{dz}{\sqrt{(1+z^2)(1+k^2 z^2)^3}}, \tag{A.1}$$

$$B = \frac{3P(\theta_1 + \theta_2)}{2b^3} I_2(k), \quad I_2(k) = \int_0^\infty \frac{dz}{\sqrt{(1+z^2)(1+k^{-2} z^2)^3}}, \tag{A.2}$$

$$\delta = \frac{3P(\theta_1 + \theta_2)}{2a} F_1(k), \quad F_1(k) = F(k_1) = \int_0^{\frac{\pi}{2}} \frac{d\varphi}{\sqrt{1 - k_1^2 \sin^2 \varphi}}, \tag{A.3}$$

where $k = \frac{b}{a} < 1$, $k_1^2 = 1 - k^2$ and $\theta_i = \frac{1 - \upsilon_i^2}{\pi E_i}$, $i = 1, 2$.

From Eqs. (A.1) and (A.2) we obtain

$$AB = \frac{9P^2(\theta_1 + \theta_2)^2 I_1(k) I_2(k)}{4a^3b^3} \qquad (A.4)$$

Let us represent the relation (A.4) by the form

$$(AB)^{1/4} = \frac{\sqrt{3}P^{1/2}(\theta_1 + \theta_2)^{1/2}(I_1(k)I_2(k))^{1/4}}{\sqrt{2}\, k^{3/4} a^{3/2}} \qquad (A.5)$$

From Eq. (A.3) it results

$$a^{3/2} = \left(\frac{3P(\theta_1 + \theta_2)}{2\delta} F_1(k)\right)^{3/2} \qquad (A.6)$$

It follows that the force relation is given by

$$P = \frac{2(I_1(k)I_2(k))^{1/4} \delta^{3/2}}{3(\theta_1 + \theta_2) k^{3/4} (F_1(k))^{3/2} (AB)^{1/4}} \qquad (A.7)$$

**Figure captions**

Fig. 1. Contact cone / plane surface.

Fig. 2. (a) Contact between two bodies of an arbitrary form, (b) Domain of intersection $\Gamma$.

Fig. 3. Evolution of the function $f_1(k,\xi)$.

Fig. 4. Contact between a plane surface and (a) a spherical surface, (b) a conical surface and (c) a pyramidal surface.

Fig. 5. Relaxation test.

Fig. 6. Comparison of the relaxation curve and the reconstitution curve with four characteristic times.

Fig. 7. Experimental device for a contact between a rubber block and (a) a sphere, (b) a cone, (c) a pyramid.

Fig. 8. Evolution of the force upon the interpenetration for the sphere contact (a), the cone contact (b) and the pyramid contact (c).

Fig. 9. Comparison between viscoelastic experimental results and analytic results for the spherical case (a), the conical case (b) and the pyramidal case (c).

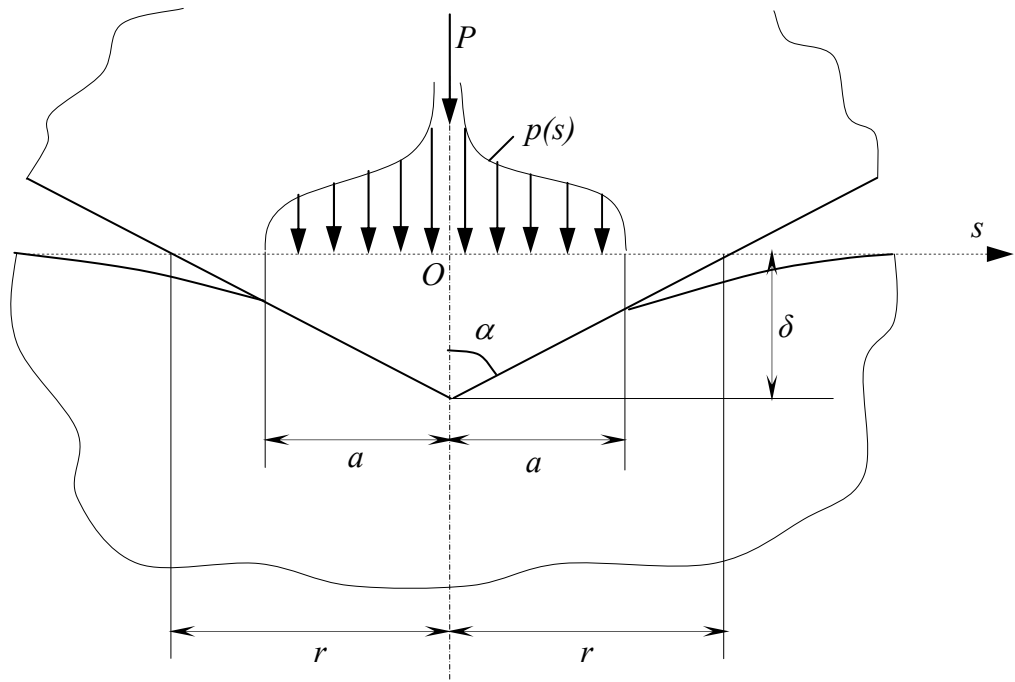

Fig. 1. Contact cone / plane surface.

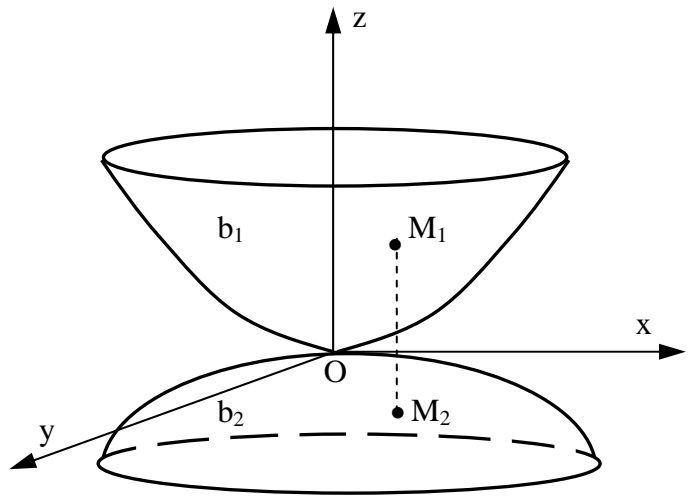

(a)

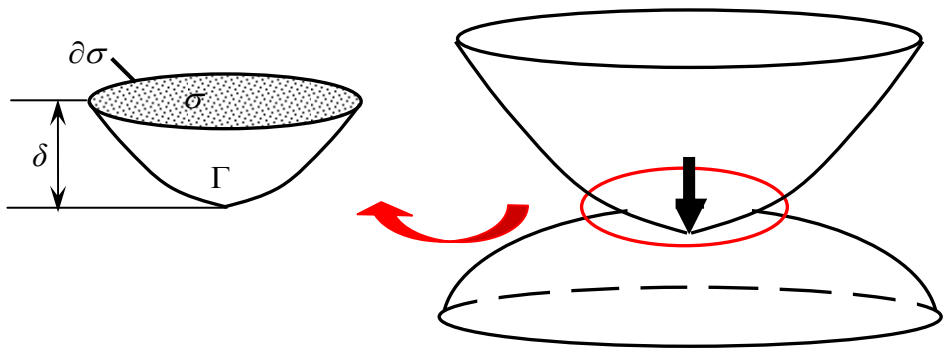

(b)

Fig. 2. (a) Contact between two bodies of an arbitrary form, (b) Domain of intersection $\Gamma$.

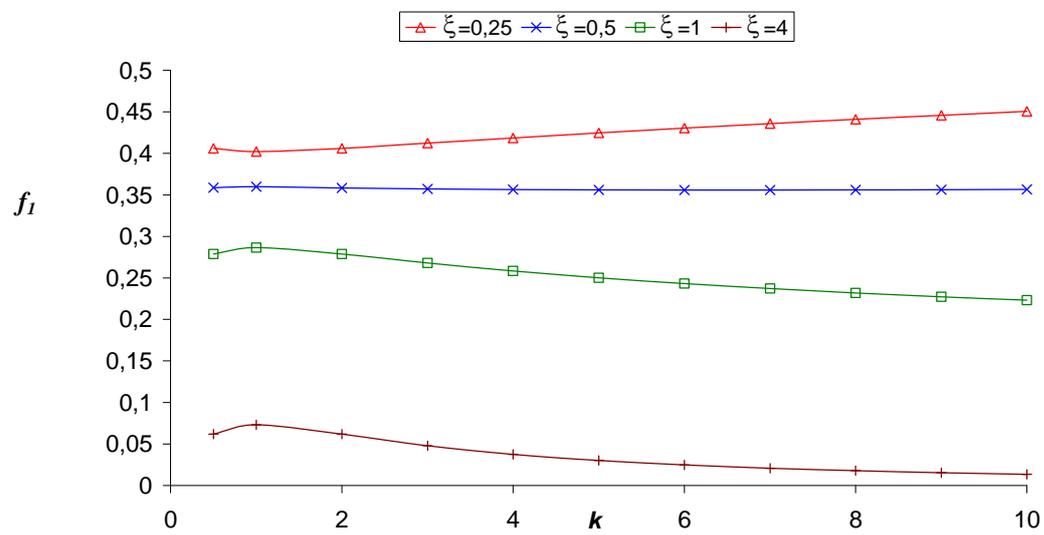

Fig. 3. Evolution of the function $f_1(k,\xi)$.

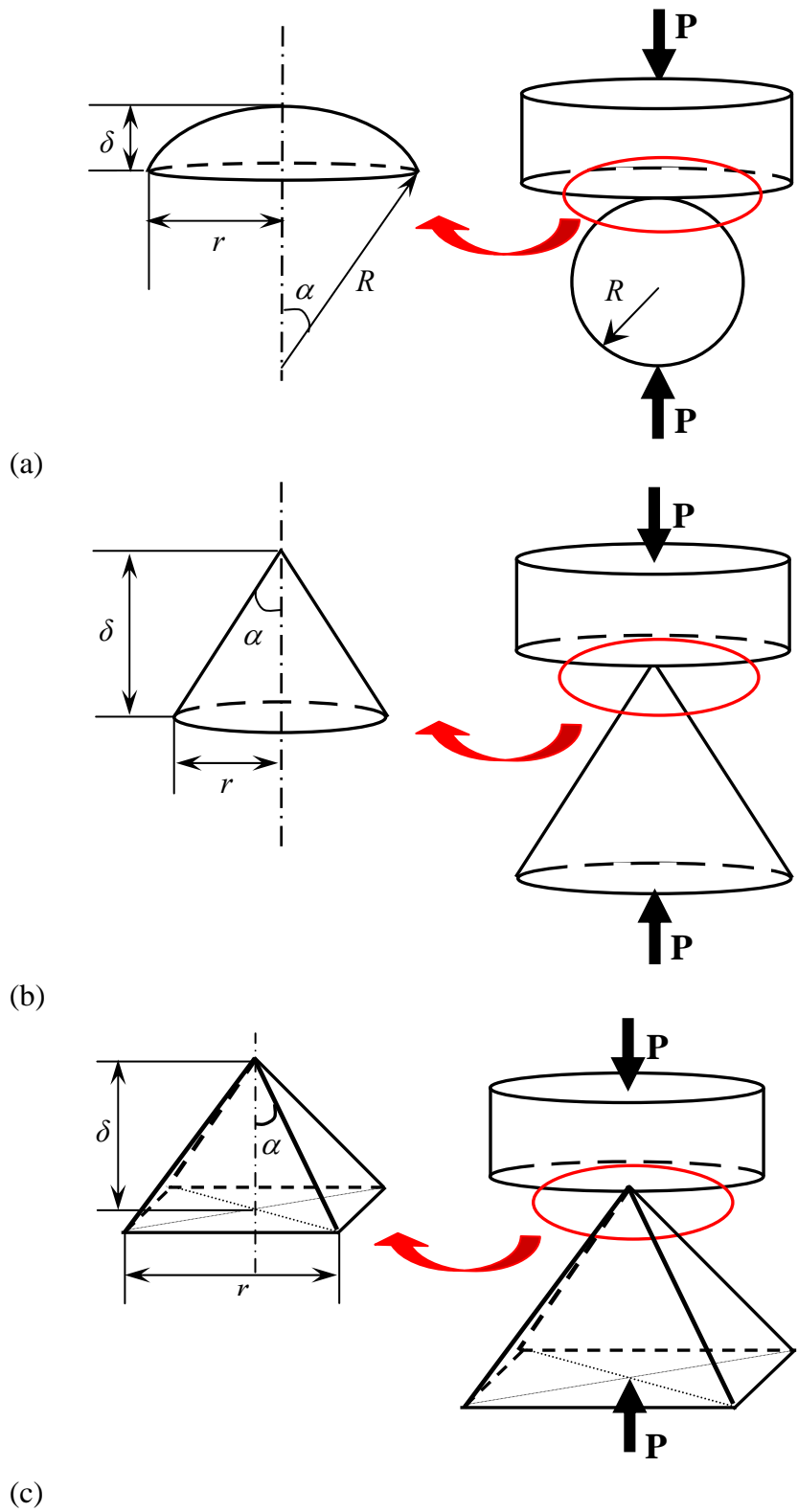

Fig. 4. Contact between a plane surface and (a) a spherical surface, (b) a conical surface and (c) a pyramidal surface.

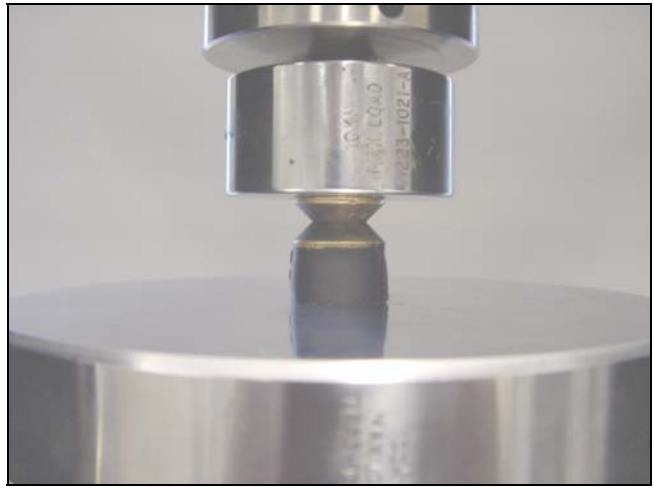

Fig. 5. Relaxation test.

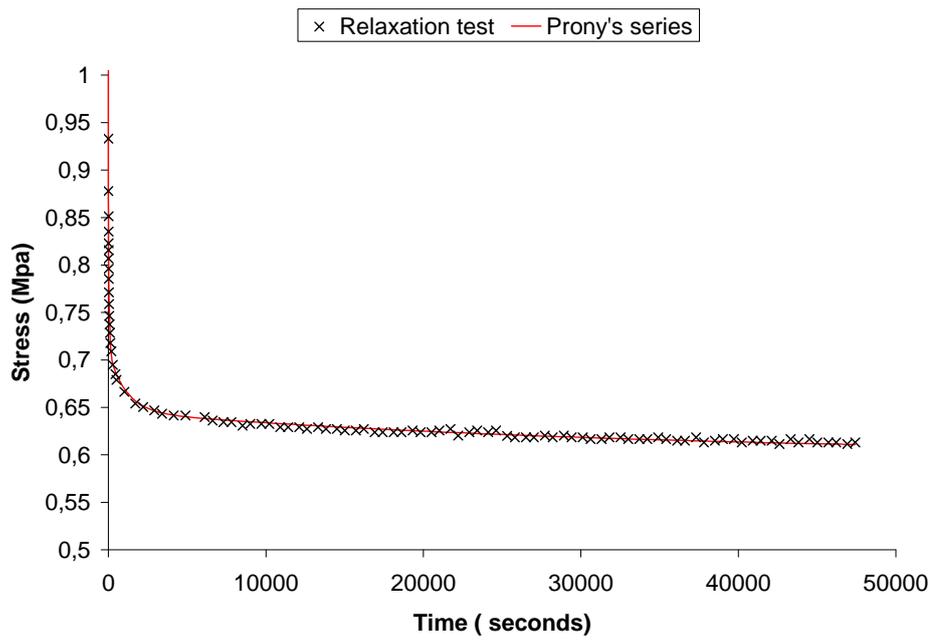

Fig. 6. Comparison of the relaxation curve and the reconstitution curve with four characteristic times.

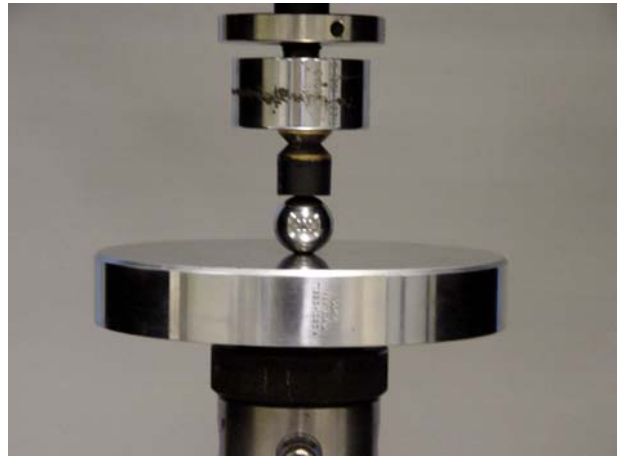

(a)

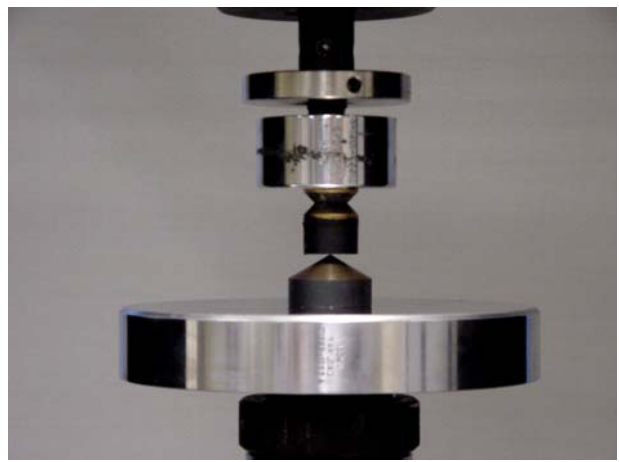

(b)

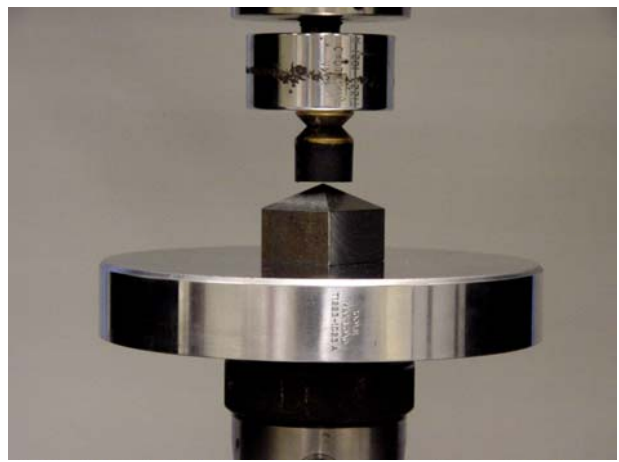

(c)

Fig. 7. Experimental device for a contact between a rubber block and (a) a sphere, (b) a cone, (c) a pyramid .

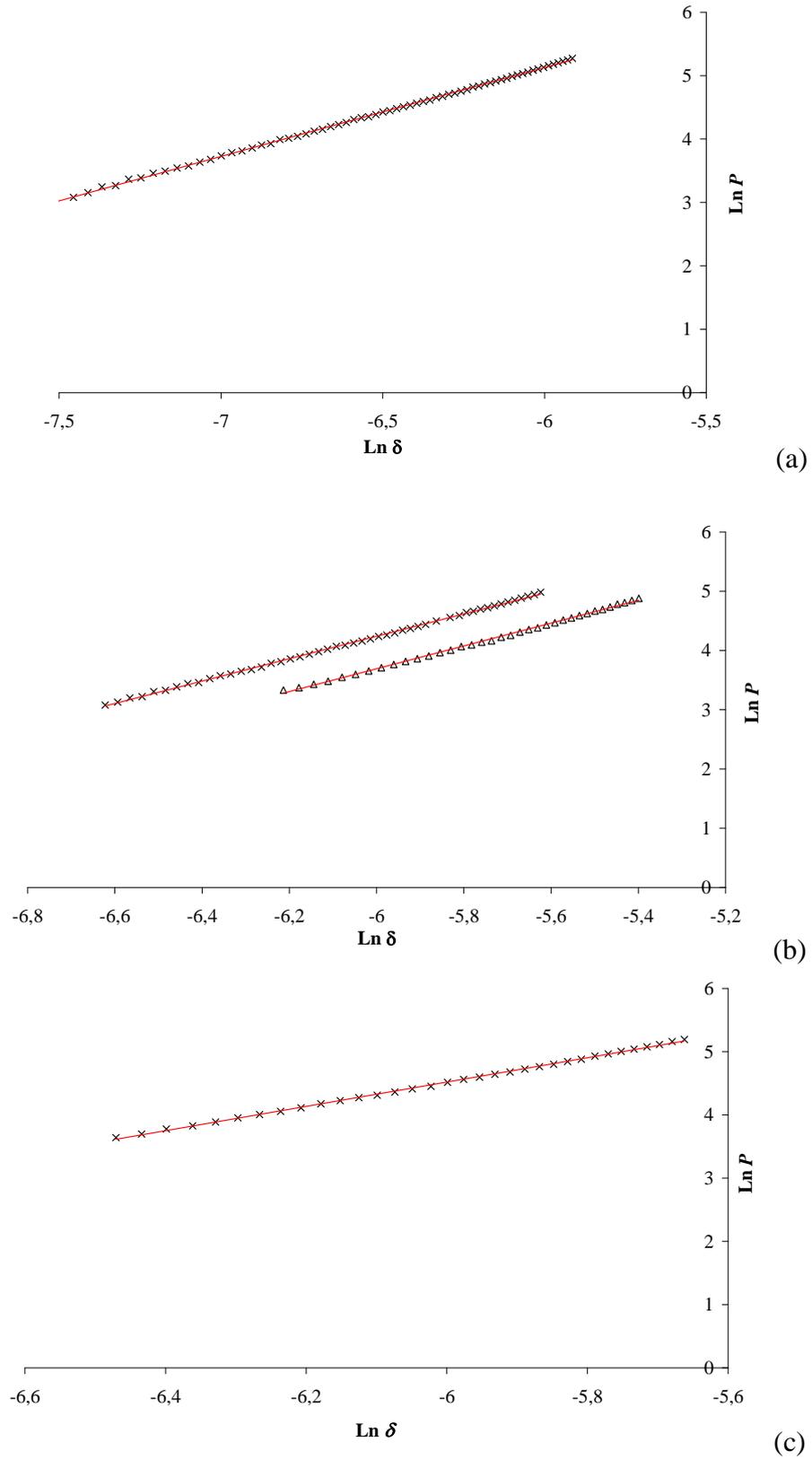

Fig. 8. Evolution of the force upon the interpenetration for the sphere contact (a), the cone contact (b) and the pyramid contact (c).

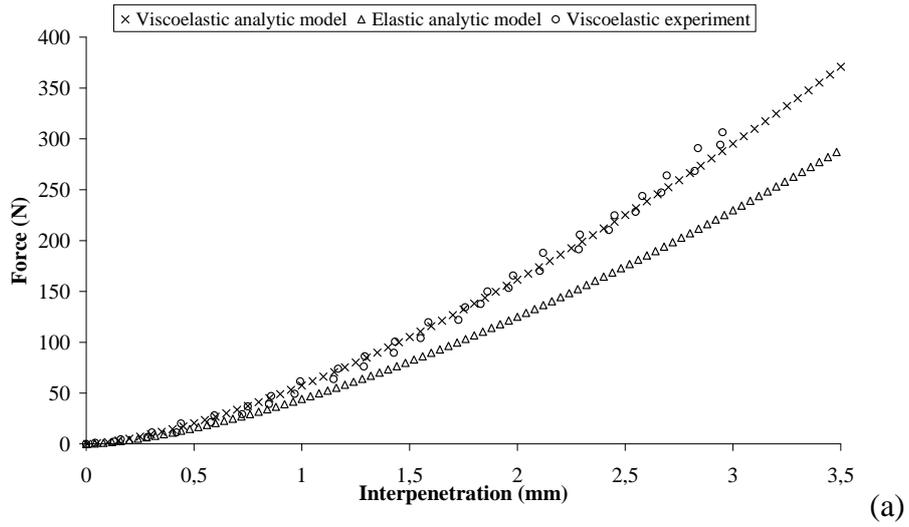

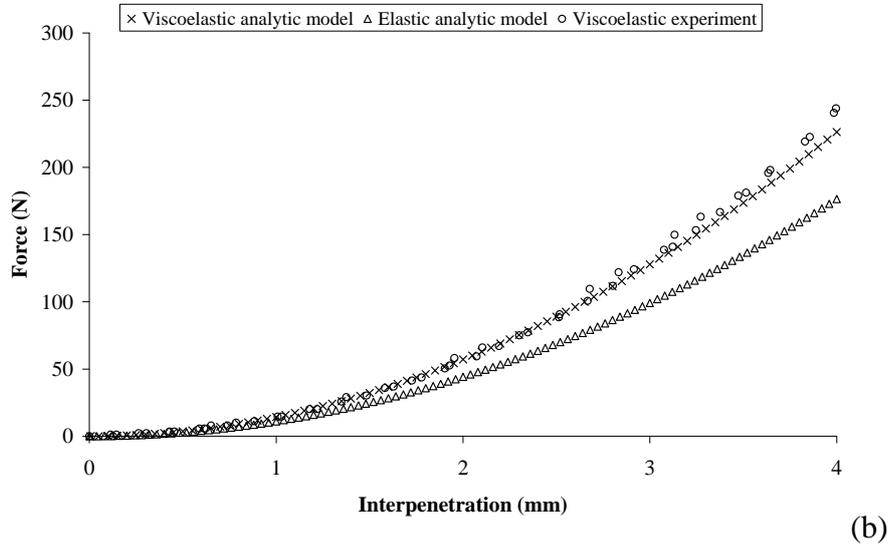

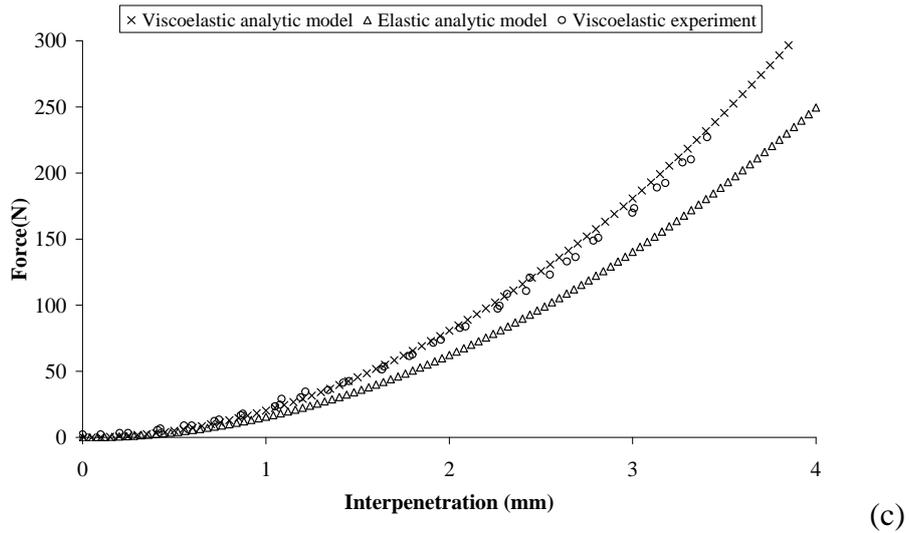

Fig. 9. Comparison between viscoelastic experimental results and analytic results for the spherical case (a), the conical case (b) and the pyramidal case (c).

Table 1

Parameters $\gamma$ and $Q$ for the various forms of contact surfaces.

| Shapes | *Spherical* | *Conical* | *Pyramidal* |
|---|---|---|---|
| $\gamma$ | 3/2 | 2 | 2 |
| $Q$ | $\dfrac{8}{3}\sqrt{R}$ | $\dfrac{4}{\pi}tg\,\alpha$ | $c_p \dfrac{32}{45}\sqrt{2}\pi^{3/4} tg\,\alpha$ |

Table 2

Characteristic times and amplitudes for the rubber block.

| *i* | *1* | *2* | *3* | *4* |
|---|---|---|---|---|
| $\tau_i$ *(second)* | *33333.33* | *1110.00* | *64.94* | *3.37* |
| $A_i$ *(Mpa)* | *0.58* | *0.79* | *1.56* | *2.29* |

Table 3

Identification of the coefficients $c_s$, $c_c$ et $c_p$.

| Shape | Spherical | Conical | | Pyramidal |
|---|---|---|---|---|
| C | $c_s$ | $c_c$ | | $c_p$ |
| **Experimentally** | *0.34* | **45°** | **60°** | 0.53 |
| | | *0.43* | *0.45* | |
| **Analytically** | *0.36* | 0.45 | | - |